\documentclass[onecolumn,authoryear]{article}

\usepackage{amsmath,amssymb,amsfonts,amsthm,graphicx}
\usepackage{txfonts}
\usepackage{helvet}
\usepackage[utf8x]{inputenc} %%%%%%agregar x
\usepackage{mathrsfs}
\usepackage{lmodern}
\usepackage[T1]{fontenc}
\usepackage{color}
\usepackage{authblk}
\usepackage{natbib}
\usepackage[hidelinks]{hyperref}

\newtheorem{definition}{Definition}
\theoremstyle{definition}

\newtheorem{theorem}{Theorem}

\newtheorem{proposition}{Proposition}

\def\begequarr{\begin{eqnarray}}
\def\endequarr{\end{eqnarray}}
\def\begequarrs{\begin{eqnarray*}}
\def\endequarrs{\end{eqnarray*}}
\def\begarr{\begin{array}}
\def\endarr{\end{array}}
\def\begequ{\begin{equation}}
\def\endequ{\end{equation}}
\def\lab{\label}
\def\begdes{\begin{description}}
\def\enddes{\end{description}}
\def\begenu{\begin{enumerate}}
\def\begite{\begin{itemize}}
\def\endite{\end{itemize}}
\def\endenu{\end{enumerate}}

\def\lef[{\left[\begin{array}}
\def\rig]{\end{array}\right]}

\def\begcen{\begin{center}}
\def\endcen{\end{center}}
\def\begrem{\begin{remark}\rm}
\def\endrem{\end{remark}}

\def\L2e{\call_{2e}}
\def\bul{\noindent $\bullet\;\;$}
\def\rea{\mathbb{R}}

\def\l2{{\mathcal L}_2}
\def\l2e{\call_{2e}}
\def\bul{\noindent $\bullet\;\;$}
\def\rea{\mathbb{R}}

\def\calm{{\mathcal M}}

\def\cale{{\mathcal E}}

\def\call{{\mathcal L}}

\def\call{{\mathcal L}}

\def\begmat#1{\begin{bmatrix}#1\end{bmatrix}}

\AtBeginEnvironment{array}{\setlength{\arraycolsep}{1.2pt}}
\AtBeginEnvironment{eqnarray}{\setlength{\arraycolsep}{1.2pt}}

\DeclareMathAlphabet{\mathpzc}{OT1}{pzc}{m}{it}

\newcommand{\PAR}[2]{\frac{\partial{#1}}{\partial{#2}}}           %% Partial derivatives
\newcommand{\MAP}[3]{{#1}:{#2}\to\mathbb{R}^{#3}}      %% Mappings
\newcommand{\col}{\mbox{col}}

\def\rea{\mathbb{R}}

\begin{document}

\title{Introduction to Passivity-based Control}

\author[1]{Pablo Borja}%
\author[2]{Romeo Ortega}%

%\author[1,2]{Third Author}%

\affil[1]{University of Plymouth, Plymouth, United Kingdom.}
\affil[2]{Mexico Autonomous Institute of Technology (ITAM), Mexico City, Mexico.}
\date{}

\pagestyle{empty}

\begin{center}
 \textbf{\Huge Introduction to Passivity-based Control}\\[2cm]
 {\Large Pablo Borja* and Romeo Ortega**}\\[1cm]
 *University of Plymouth, Plymouth, United Kingdom\\
 **Mexico Autonomous Institute of Technology (ITAM), Mexico City, Mexico.
\end{center}

\vspace*{3cm}

\noindent\textbf{Disclaimer / Notice:} This is a preprint of an encyclopedia chapter accepted for 
publication in Encyclopedia of Systems and Control Engineering -- Volume 1, edited by Sarah 
Spurgeon, published by Elsevier. This version has not undergone final publisher proofreading or 
typesetting. The version of Record is available online via DOI:\\[0.15cm]
\url{https://doi.org/10.1016/B978-0-443-14081-5.00031-3}

\section*{Abstract}
Passivity-based control (PBC) is a nonlinear control design framework that has proven adequate for 
controlling a wide range of systems, especially physical ones. Their main ingredients are physical 
quantities such as energy and dissipation, making the control design more intuitive and endowing 
the 
controllers with a physical interpretation. In contrast to other, mathematically-based nonlinear 
control approaches, the energy-based viewpoint and physical intuition of PBC often make this 
strategy more robust and energy efficient.\\
This chapter provides an overview of PBC, revisiting the basic aspects of this powerful nonlinear 
control framework and the most common PBC approaches.\\[0.15cm]
\textbf{Keywords:} Nonlinear systems; Lyapunov stability; Passivity; Energy; Dissipation; 
Stabilization; Storage 
function; Physical systems; Energy shaping; Damping injection; Interconnection; Feedback.

\newpage

\maketitle

\textbf{Nomenclature}\\
\begin{equation*}
\begin{array}{ll}
\text{CbI} &\text{Control by interconnection}\\
\text{EL} &\text{Euler-Lagrange}\\
\text{IDA} &\text{Interconnection and damping assignment}\\
\text{PBC} &\text{Passivity-based control}\\
\text{PDE} &\text{Partial differential equation}\\
\text{pH} &\text{Port-Hamiltonian}\\
\text{PID} &\text{Proportional-integral-derivative}
\end{array}
\end{equation*}

% \end{frontmatter}
%%%%%%%%%%%%%%%%%%%%%%%
%%%%%%%%%%%%%%%%%%%%%%%
%  \section*{Key Points}
% %%%%%%%%%%%%%%%%%%%%%%%
% %%%%%%%%%%%%%%%%%%%%%%%
% 
% The key points of this chapter are:
% \begin{itemize}
%  \item Revisit the concepts of passivity and passive systems in the context of input-affine 
% nonlinear systems.
%  \item Describe the main idea and basic elements of passivity-based control.
%  \item Discuss common challenges in the passivity-based control design.
%  \item Provide an overview of the most relevant passivity-based control approaches.
% \end{itemize}

%%%%%%%%%%%%%%%%%%%%%%%
%%%%%%%%%%%%%%%%%%%%%%%
\section{Introduction}
%%%%%%%%%%%%%%%%%%%%%%%
%%%%%%%%%%%%%%%%%%%%%%%
Since the term passivity-based control (PBC) was coined in \cite{ortega1989adaptive}, this 
nonlinear control design framework has proven suitable for controlling many nonlinear systems. The 
key idea in PBC is that the controller is an energy-transforming and dissipating device that, once 
interconnected to the plant, ensures that the dissipation and energy of the overall system have a 
desired shape. Hence, PBC is a natural way to control physical systems, where energy plays a central 
role in determining their behavior. In particular, PBC methods have proven successful in 
controlling systems modeled using energy-based approaches, e.g., the port-Hamiltonian (pH) 
framework \citep{GEObook,VANJEL} and the Euler-Lagrange (EL) formalism \citep{goldstein,ORTbook,SPONGVID}. 
Another advantage of the energy-based perspective of PBC with respect to the traditional 
signal-processing viewpoint of other nonlinear control techniques is that physical 
phenomena---typically nonlinear---can be seamlessly accounted for in the control design process, 
endowing the resulting controllers with a physical interpretation and making them more intuitive. 
Moreover, passivity-based controllers do not rely on canceling nonlinearities to fit into a desired 
mathematical structure, e.g., linear models. Hence, these controllers are often more robust and 
energy-efficient than other nonlinear controllers.

This chapter revisits basic concepts related to PBC, such as passive systems, energy shaping, and 
damping injection. Then, it provides an overview, including challenges and main features, of three 
of the most relevant PBC approaches: control by interconnection (CbI), 
proportional-integral-derivative (PID) PBC, and interconnection and damping assignment (IDA) PBC.

\subsection*{Notation}
The symbol $I$ denotes the identity matrix. The symbol $\mathbf{0}$ is reserved for vectors and 
matrices whose entries are zeros. The dimensions of $I$ and $\mathbf{0}$ follow from the 
context. Given $x(t)\in\rea^{n}$, $t\in\rea$, and the mapping $\MAP{f}{\rea^n}{n}$, we adopt 
the notation 
$$\nabla f(x) = \left[\PAR{f(x)}{x_{1}} \; \dots \;\PAR{f(x)}{x_{n}}\right]^{\top}, 
\qquad \dot{x}=\dfrac{\mathrm{d}x(t)}{\mathrm{d}t}.$$
Given the (square) symmetric matrix $A\in\rea^{n\times n}$, the notation $A\succ 0$ indicates that 
$A$ is positive definite. Likewise, $A\succeq 0$, $A\prec 0$, and $A\preceq 0$ denote that $A$ is 
positive semi-definite, negative definite, and negative semi-definite, respectively. Given 
$x\in\rea^{n}$ and a positive (semi-)definite matrix $A\in\rea^{n\times n}$, $\lVert x \rVert$ and 
$\lVert x \rVert_{A}$ represent the Euclidean norm and a weighted Euclidean norm, respectively, 
i.e.,
$$\lVert x \rVert:=\displaystyle\sqrt{x^{\top}x}, \qquad  \lVert x 
\rVert_{A}:=\displaystyle\sqrt{x^{\top}Ax} $$
Given the distinguished vector $x^{\star}\in\rea^{n}$ and the mapping $\MAP{f}{\rea^n}{n}$, we 
define 
$$f^{\star}:=f(x^{\star}), \qquad \left( \nabla f \right)^{\star}:=\nabla f(x^{\star}).$$
The symbol $\col(\cdot)$ is a compact form to express a column vector, e.g., $\col(x_{1},x_{2}) = [ x_{1} \; x_{2}]^{\top}$.
%%%%%%%%%%%%%%%%%%%%%%%
%%%%%%%%%%%%%%%%%%%%%%%
\section{Passivity and Passive Systems}
%%%%%%%%%%%%%%%%%%%%%%%
%%%%%%%%%%%%%%%%%%%%%%%
The concept of passivity has its roots in electrical circuit theory \citep{moylan2014}. However, it
has been extended to general dynamical systems, where, loosely speaking, a system is passive if it 
cannot generate energy. Here, we focus on systems that are described by the following state-space 
representation:
\begin{equation}\label{sys}
\begin{array}{rcl}
  \dot{x} &=& f(x(t))+g(x(t))u(t) \\
  y(t) &=& h(x(t))+j(x(t))u(t),
\end{array}
\end{equation}
where $t\in\rea_{\geq0}$ represents time; the states of the system are given by 
$x:\rea_{\geq0}\to\mathcal{X}$, with $\mathcal{X}$ a differentiable $n$-dimensional 
manifold;\footnote{Often in the literature, $\mathcal{X}\subseteq\rea^{n}$.} 
$u:\rea_{\geq0}\to\rea^{m}$ denotes the input to the system, with $m\leq n$; 
$y:\rea_{\geq0}\to\rea^{m}$
is the system's output; $\MAP{f}{\mathcal{X}}{n}$; $\MAP{g}{\mathcal{X}}{n\times m}$; $\MAP{h}{\mathcal{X}}{m}$; $\MAP{j}{\mathcal{X}}{m\times m}$. The inner product between the input and output, i.e., $u^{\top}(t)y(t)$, is called the supply rate and has power units. Moreover, we assume that the solution $x(t)$ of \eqref{sys} is unique, $u(t)$ and $y(t)$ are bounded, and $\int_{0}^{t}u^{\top}(\tau)y(\tau) d\tau$ is well-defined.

The following definition formalizes the concept of passive systems---for systems of the form \eqref{sys}.
\begin{definition}
Let $\Sigma: u \to y$ denote a system described by \eqref{sys}. $\Sigma$ is said to be 
passive if there exists $S:\mathcal{X}\to\rea_{\geq0}$ satisfying
\begin{equation}
 S(x(t))\leq S(x(0))+\displaystyle\int_{0}^{t}u^{\top}(\tau)y(\tau) d\tau. \label{disin}
\end{equation}
Moreover, $\Sigma$ is said to be:
\begin{itemize}
 \item Input strictly passive if there exists $\delta>0$ such that
\begin{equation*}
 S(x(t))\leq S(x(0))+\displaystyle\int_{0}^{t}\left[u^{\top}(\tau)y(\tau)-\delta\lVert u(\tau) 
\rVert^{2} \right]d\tau.
\end{equation*}
 \item Output strictly passive if there exists $\varepsilon>0$ such that
\begin{equation*}
 S(x(t))\leq S(x(0))+\displaystyle\int_{0}^{t}\left[u^{\top}(\tau)y(\tau)-\varepsilon\lVert y(\tau) 
\rVert^{2} \right]d\tau.
\end{equation*}
\end{itemize}

\end{definition}
The inequality \eqref{disin} is referred to as the dissipation inequality, and any function 
$S(x(t))$ satisfying such an inequality is known as a storage function \citep{WIL}. Furthermore, if 
\eqref{disin} holds, $y(t)$ is said to be a {passive output} of \eqref{sys}. The supply rate 
denotes 
the power flowing into the system. Hence, storage functions are associated with energy---physical 
or 
virtual. A consequence of  \eqref{disin}, and the fact that $S(x(t))\geq 0$, is the inequality
$$
-\int_{0}^{t}u^{\top}(\tau)y(\tau) d\tau \leq  S(x(0)),
$$
that, in view of the minus sign in the first term, can be interpreted as: ``we cannot extract from a 
passive  system more energy than the one originally stored.''  Assuming differentiability of 
$S(x(t))$, the inequality
\eqref{disin} can be expressed as
\begin{equation}
 \dot{S} \leq u^{\top}(t)y(t). \label{ddisin}
\end{equation}
The inequality \eqref{ddisin} is often referred to as the differential dissipation inequality. Such 
an inequality implies that the power that can be extracted from the system at $t$ cannot be
greater than the power injected into the system. The following theorem---known as 
Hill-Moylan's theorem \citep{hill76}---establishes conditions to verify if a system described by 
\eqref{sys} is passive.
\begin{theorem}\label{th:hill}\em
 A system of the form \eqref{sys} is passive with a differentiable storage function $S(x(t))$ if and 
only if
\begin{equation*}
 \begin{array}{rcl}
  \nabla^{\top}S(x(t))f(x(t)) &=& -\lVert\ell(x(t))\rVert^{2} \leq 0 \\
  h(x(t)) &=& g^{\top}(x(t))\nabla S(x(t)) + 2w^{\top}(x(t))\ell(x(t)) \\
  j(x(t)) &=& w^{\top}(x(t))w(x(t)) + D(x(t))
 \end{array}
\end{equation*}
for some $\MAP{\ell}{\mathcal{X}}{q}$, with $q\in\mathbb{N}_+$, $\MAP{w}{\mathcal{X}}{q\times {m}}$, 
and skew-symmetric $\MAP{D}{\mathcal{X}}{{m}\times {m} }$.
 \end{theorem}

\begin{figure}[htp]
\centering
\includegraphics[width=.5\textwidth]{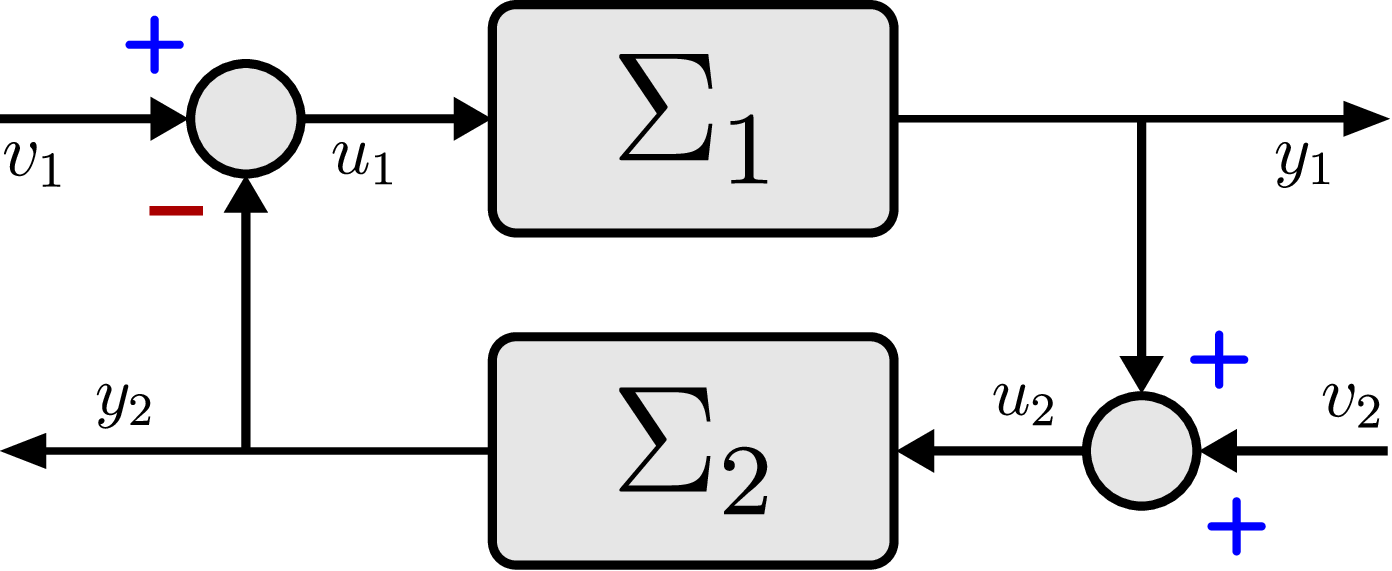}
	\caption{Standard negative feedback interconnection of two systems. The blocks $\Sigma_{1}$ and 
$\Sigma_{2}$ represent dynamical systems. Similarly, the signals $u_{1}, u_{2}$ represent the 
inputs, $y_{1}, y_{2}$ denote the outputs, and $v_{1}, v_{2}$ represent external signals.}
	\label{fig:feed}
\end{figure}
One of the most celebrated results from passivity theory is given by the passivity theorem 
\citep{KHA,VAN, PIDbook}, which, informally speaking, states that the negative feedback 
interconnection of two passive systems $\Sigma_{1}$ and $\Sigma_{2}$---see Figure 
\ref{fig:feed}---yields another passive system. To formalize this idea, we first introduce the 
following definition:
\begin{definition}
 A system described by \eqref{sys} has $\mathcal{L}_{2}$-gain less than or equal to $\rho>0$ 
if\footnote{The symbol $\lVert (\cdot) \rVert_{2}$ denotes the $\mathcal{L}_{2}$-norm---see 
\cite{VAN}.}
 \begin{equation*}
  \lVert y(t) \rVert^{2}_{2}\leq \rho\lVert u(t) \rVert^{2}_{2} + \beta, \qquad \beta\geq 0.
 \end{equation*}
\end{definition}
Notably, if \eqref{sys} is output strictly passive, then it has $\mathcal{L}_{2}$-gain less than or 
equal to $\dfrac{1}{\varepsilon}$ \citep{KHA,VAN}. Accordingly, a general version of the passivity 
theorem is given by the following proposition.
\begin{proposition}\label{pro:passth}\em
 Consider the negative feedback interconnection depicted in Figure \ref{fig:feed}, where the 
subsystems $\Sigma_{1}$ and $\Sigma_{2}$ satisfy
 \begin{equation*}
  S_{i}(x_{i}(t))\leq 
S_{i}(x_{i}(0))+\displaystyle\int_{0}^{t}\left[u_{i}^{\top}(\tau)y_{i}(\tau)-\delta_{i}\lVert 
u_{i}(\tau) \rVert^{2}-\varepsilon_{i}\lVert y_{i}(\tau) \rVert^{2} \right]d\tau, \qquad  
i\in\{1,2\},
 \end{equation*}
for some $\varepsilon_i,\delta_i \in \rea$. The interconnected system $\col(v_{1},v_{2})\to 
\col(y_{1},y_{2})$ has finite $\mathcal{L}_{2}$-gain if
\begin{equation*}
 \varepsilon_{1}+\delta_{1}>0, \qquad \varepsilon_{2}+\delta_{2}>0.
\end{equation*}
\end{proposition}
A consequence of Proposition \ref{pro:passth} is that the lack of passivity is compensated by an 
excess of passivity, which is equivalent to say that energy generation is compensated by 
dissipation.

This chapter focuses on input-affine nonlinear systems described by \eqref{sys}. However, the concept of passive systems can be extended to more general dynamical systems. To this end, we refer the reader to \cite{moylan2014, VAN}.\\[0.1cm]
\textbf{Caveat:} In the subsequent sections,  to simplify the notation, the argument $t$ is removed from all signals.

%%%%%%%%%%%%%%%%%%%%%%%
\section{Passivity-based Control, Energy Shaping, and Damping Injection}
%%%%%%%%%%%%%%%%%%%%%%%
%%%%%%%%%%%%%%%%%%%%%%%
The key idea behind PBC in state regulation tasks is to exploit the passivity property of a system to 
design the input $u$ such that the desired configuration $x^{\star} \in \mathcal{X}$ is a stable 
equilibrium for the regulated system. The first step towards this end is to define the set of 
equilibria that can be 
assigned by the control input. In particular, considering systems of the described by \eqref{sys}, 
the set of assignable equilibria is given by
\begin{equation*}
\mathcal{E}:=\left\lbrace x\in\mathcal{X} \mid g^{\perp}(x)f(x) = \mathbf{0} \right\rbrace,
\end{equation*}
where $g^{\perp}(x)$ denotes the left-annihilator of $g(x)$, which is a full-rank matrix with $n-m$ rows and $n$ columns such that $g^{\perp}(x)g(x) = \mathbf{0}$.

Customarily, the design of a passivity-based controller consists of two steps, namely:
\begin{itemize}
 \item {Energy shaping:} In this step, the storage function for the regulated system is designed. This function---which is going to be used as Lyapunov function for the stability analysis---must have a strict minimum at $x^{\star}$.
 \item {Damping injection:} Once the desired storage function has been assigned to the regulated system, damping is injected to guarantee the system loses energy until the states converge to the minimum of the storage function, i.e., the desired equilibrium.
\end{itemize}
The energy shaping step ensures the existence of a function $S_{\tt d}(x)$ that is locally positive 
definite at the desired equilibrium, i.e.,
\begin{equation}
 S_{\tt d}^\star = 0 \qquad \text{and} \qquad S_{\tt d}(x) > 0; \; \forall 
x\in\mathcal{D}-\{x^{\star}\}, \label{posdef}
\end{equation}
where $\mathcal{D}\subseteq\mathcal{X}$ is an open set containing $x^{\star}$. Additionally, because of the passivity property,
\begin{equation}
 \dot{S}_{\tt d} \leq u_{\tt di}^{\top}y, \label{clin}
\end{equation}
where $u_{\tt di}$ denotes the controller part corresponding to the damping injection step. Hence, 
by selecting $u_{\tt di} = -K_{\tt di}y$, with $K_{\tt di}\in\rea^{n\times n}$ positive definite, 
\eqref{clin} reduces to
\begin{equation}
 \dot{S}_{\tt d} \leq - \lVert y \rVert^{2}_{K_{\tt di}} \leq 0. \label{noninc}
\end{equation}
Note that \eqref{posdef} and \eqref{noninc} imply that $S_{\tt d}(x)$ qualifies as a Lyapunov 
function, and $x^{\star}$ is a locally stable equilibrium for the closed-loop system. Additionally, 
if the detectability condition
\begin{equation*}
  y(t) \equiv \mathbf{0} \implies  x(t) \to x^{\star}
\end{equation*}
holds, the desired equilibrium is locally asymptotically stable. Similarly, if $S_{\tt d}(x)$ is 
radially 
unbounded, the stability properties are global. For further details on Lyapunov stability, we refer 
the reader to \cite{vid,KHA,VAN}.

From a physical perspective, the passivity property amounts to reformulating energy balancing. 
Therefore, identifying storage functions and passive outputs is straightforward in many physical 
systems. This makes PBC a powerful approach to control a wide variety of physical systems, as 
illustrated by the vast literature reporting successful PBC implementations in engineering 
applications, e.g., \cite{ORTbook,GEObook,PIDbook}. Moreover, since the 
concept of energy is paramount in PBC, its synergy with the pH 
framework and the EL formalism is remarkable.

The different PBC strategies can be divided into two main groups \citep{ORTcsm}:

\begin{itemize}
    \item[{\bf (i)}] Approaches where only the closed-loop structure is fixed. In this case, the 
admissible interconnection pattern, dissipation, and storage function are characterized by a 
partial differential equation (PDE). 
Some examples of these approaches are IDA-PBC 
\citep{IDAaut,IDAsurvey}, the controlled Lagrangians method \citep{blochI,blochII}, and methods 
based on coordinate transformations \citep{FUJIetaltracking}.
    \item[{\bf (ii)}] Techniques where the structure of the storage function is predetermined with 
some degrees of freedom that can be modified by the controller. In this case, the control law is 
designed to ensure that the storage function qualifies as a Lyapunov function to prove the stability of
the desired equilibrium. Some examples of these techniques are energy-balancing PBC 
\citep{ORTcsm,ORTetaltac08,CASORT} 
and PID-PBC \citep{PIDbook,gandarilla2021pid,zonetti2022pid}.
\end{itemize}

The PBC strategies encompassed in {\bf (i)} are suitable for stabilizing a broader range of plants. 
Unfortunately, their implementation is sometimes hindered by the complexity of the PDEs involved in 
the control design process. In contrast, the techniques belonging to {\bf (ii)} are suitable for 
controlling a more limited family of systems but often avoid solving PDEs, and the resulting 
controllers are endowed with a more transparent physical interpretation, making them more systematic 
and intuitive. In the subsequent sections, we provide an overview of some of the most relevant PBC 
techniques. However, due to space constraints, we do not revisit more specific  PBC 
approaches, e.g.,
Krasovskii PBC \citep{kawano,kaw23}.

%%%%%%%%%%%%%%%%%%%%%%%%%%%%%%%%%%%%%%%%%%%%%%%%
\section{Different Passive Outputs and the Dissipation Obstacle}
%%%%%%%%%%%%%%%%%%%%%%%%%%%%%%%%%%%%%%%%%%%%%%%%

In the control engineering context, the signal $u$ often corresponds to the control input to be 
designed to achieve a specific control task, for instance, stabilizing a desired equilibrium. In 
PBC, $u$ depends on the passive output $y$ to inject damping and, depending on the PBC approach, to 
shape the energy of the system. However, it is important to remark that there exist different 
passive outputs associated with the same storage function. In some PBC approaches, e.g., PID-PBC or 
CbI, identifying an adequate passive output is crucial to assigning the desired equilibrium and 
circumventing some issues, such as the dissipation obstacle. This section is devoted to identifying 
the passive outputs associated with a given storage function and illustrating the deleterious 
effects of the dissipation obstacle in PBC.
%%%%%%%%%%%%%%%%%%%%%%%%%%%%%%%%%%%%%%%%%%%%%%%%
\subsection{Parameterization of All the Passive Outputs for a Given Storage 
Function}\label{sec:par}
%%%%%%%%%%%%%%%%%%%%%%%%%%%%%%%%%%%%%%%%%%%%%%%%
Given Hill-Moylan's theorem---see Theorem \ref{th:hill}---we can deduce that all the passive 
outputs associated with the storage function $S(x)$ are parameterized as follows:
\begin{equation}\label{ypar}
y = g^{\top}(x)\nabla S(x) + 2w^{\top}(x)\ell(x)+\left[ w^{\top}(x)w(x) + D(x) \right]u.
\end{equation}
Note that the so-called feedthrough term $j(x)$ determines the relative degree of the system. In particular, $j(x)=\mathbf{0}$ implies that $w(x)=\mathbf{0}$ and $D(x)=\mathbf{0}$. Consequently, the passive output is given by
\begin{equation}\label{yn}
 y_{\tt n}:= g^{\top}(x)\nabla S(x).
\end{equation}
This particular passive output is commonly known as the natural output---hence the subscript 
${\tt n}$ to differentiate it from the general case described in \eqref{ypar}. The selection of 
$w(x)$ and $D(x)$ in \eqref{ypar} determines if the passive output is integrable, its relative 
degree and if \eqref{sys} is affected by the dissipation obstacle---see Section \ref{sec:obstacle}. 
All the mentioned features affect the PBC design and can even determine whether a PBC approach is 
suitable for controlling a specific plant.

%%%%%%%%%%%%%%%%%%%%%%%
%%%%%%%%%%%%%%%%%%%%%%%
\subsection{Passivity-based Control and the Dissipation Obstacle}
\label{sec:obstacle}
%%%%%%%%%%%%%%%%%%%%%%%%%%%%
%
Passive systems with relative degree one are affected by a phenomenon related to the 
dissipation; namely, there exists a steady state only if the energy extracted from the controller 
is zero at the equilibrium. This phenomenon is known as the dissipation obstacle in the PBC 
jargon and affects many physical systems. An illustrative example of this occurs in electrical 
circuits with capacitors in parallel---or inductors in series---with resistors. Such capacitors and 
inductors are leaky energy-storing elements operating at non-zero equilibria. Notably, the 
dissipation obstacle is absent in position regulation of mechanical systems because dissipation 
(due to friction) is zero when the system is not moving.

To mathematically represent the dissipation obstacle, consider a passive system of the form 
\eqref{sys} with storage function $S(x)$. Hence,
\begin{equation}
\label{cha2_powbalequ}
\dot S = \nabla^\top S(x)f(x) + \nabla^\top S(x) g(x)u.
\end{equation}
If the passive output has relative degree one---i.e., $y=y_{\tt n}$, with $y_{\tt n}$ defined as in 
\eqref{yn}---the equality \eqref{cha2_powbalequ} can be rewritten as
\begin{equation*}
 \dot{S} = \nabla^\top S(x)f(x) + y_{\tt n}^{\top}u,
\end{equation*} 
where $y_{\tt n}^\top u$ represents the supplied power and $\nabla^\top S(x)f(x)$ is 
the natural dissipation of the system. In passivity theory it is
said that the system \eqref{sys}---with $y=y_{\tt n}$---does not suffer from the  dissipation 
obstacle at the desired equilibrium $x^\star  \in \cale$ if
\begequ
\lab{cha2_disobs}
(\nabla^\top S)^\star f^\star  = 0.
\endequ

The origin of the dissipation obstacle is the existence of pervasive dissipation, which is 
dissipation that is present even at the equilibrium state. The PBC literature has thoroughly 
discussed this multifaceted phenomenon. In particular, it has been shown that the key energy
shaping step of PBC \citep[Proposition 2]{ORTetaltac08}, the generation of Casimir functions for CbI
\citep[Remark 7.1.9]{VAN}, and the assignment of a minimum at the desired point to the shaped
energy function \citep[Proposition 2]{MENetal} are all stymied by the dissipation obstacle. Furthermore, in \cite{TACBOR21}, the authors show
that passive systems with relative degree one have a constant equilibrium only when the 
dissipation obstacle is absent. To illustrate this, consider the system \eqref{sys} with 
$y=y_{\tt n}$ 
and the following dynamic extension:
	\begequarrs
		\dot x_{\tt c} &=& g^\top(x)\nabla S(x)\\
		\dot \xi &=& f_{\tt c}(\xi,x,x_{\tt c},u),
	\endequarrs
	where $x_{\tt c}\in\rea^{m}$ and $\xi \in \rea^{n_\xi}$.
Define the state of the augmented system as $\chi:=\col(x,x_{\tt c},\xi)$ such that
\begin{equation*}
 \dot{\chi} = F(\chi,u); \quad F(\chi,u): = \begin{bmatrix}
                                            f(x)+g(x)u \\ g^\top(x)\nabla S(x) \\ f_{\tt 
c}(\xi,x,x_{\tt c},u) 
                                           \end{bmatrix}.
\end{equation*} 
Suppose that the pair $(\chi^{\star},u^{\star})$ solves the equilibrium equation
\begin{equation}\label{eqdis}
 F(\chi^\star ,u^\star )=\mathbf{0},
\end{equation} 
which implies
\begin{equation*}
 \begmat{ \dot x  \\ \dot x_{\tt c}}=\mathbf{0}. 
\end{equation*} 
Hence,
\begin{equation}\label{dxdis}
 \dot x =\mathbf{0} \implies \dot S = 0  \iff (\nabla^{\top} S)^\star f^\star  + 
(\nabla^{\top} S)^\star g^\star u^{\star}  = 0
\end{equation} 
and
\begin{equation}\label{dxcdis}
 \dot x_{\tt c}=\mathbf{0} \iff  (\nabla^{\top} S)^\star   g^\star =\mathbf{0}.
\end{equation} 
Combining \eqref{dxdis} and \eqref{dxcdis} yields
\begin{equation*}
 \dot x =\mathbf{0} \implies \dot S = 0  \iff (\nabla^{\top} S)^\star f^\star  = 0,
\end{equation*} 
making evident that a constant solution $(\chi^{\star},u^{\star})$ to \eqref{eqdis} exists only if 
\eqref{cha2_disobs} holds, i.e., only if the system does not suffer from the dissipation obstacle.

%
%%%%%%%%%%%%%%%%%%%%%%%
\section{Control by Interconnection}\label{sec:CbI}
%%%%%%%%%%%%%%%%%%%%%%%
In CbI, the controller is a passive system interconnected to the plant via a power-preserving 
interconnection subsystem. Then, the resulting closed-loop system is passive, and its storage 
function depends on the storage functions of the plant and the controller. Accordingly, the gist of 
CbI is to design the controller system in such a way that the storage function of the plant is given 
a desired shape. To this end, the states of the plant and controller must be related to perform the 
energy-shaping process. Designing a dynamic control system, often permits 
interpreting the controller as a physical device, which is appealing from 
practical and conceptual viewpoints. The CbI approach has been extensively studied for pH 
systems---see, for instance, 
\cite{macchelli2006,ORTetaltac08,GEObook,VENVAN,VANJEL,FERMIDDON,IOHD}---where the controller 
often admits a straightforward physical interpretation, allowing us to
understand it as an energy exchange process instead of the standard signal-processing viewpoint. 
Extensions to other kinds of systems, e.g., port-thermodynamic systems \citep{pthermo}, can also be 
found in the literature. Below, we provide an overview of the traditional CbI formulation---first 
proposed in \cite{STR} and later extended in \cite{ORTcsm}.

In the simplest and most common formulation of CbI, the controller dynamics are given by
\begin{equation}\label{contcbi}
 \begin{array}{rcll}
  \dot{x}_{\tt c} &=& u_{\tt c}, & \quad x_{\tt c},u_{\tt c},y_{\tt c}\in\rea^{m}\\
  y_{\tt c} &=& \nabla S_{\tt c}(x_{\tt c}), 
 \end{array} 
\end{equation} 
where some simple computations show that $S_{\tt c}(x_{\tt c})$ is a storage function for 
\eqref{contcbi}. 
Hence, invoking the passivity theorem, the negative feedback interconnection of the system 
\eqref{sys} with the dynamical controller \eqref{contcbi} yields a new passive system---see Figure 
\ref{fig:feed}, where the plant corresponds to $\Sigma_{1}$ and the controller to $\Sigma_{2}$. To 
show this, note that, considering $v_{1}=v$ and $v_{2}=\mathbf{0}$, the proposed interconnection 
scheme implies that
\begin{equation*}
 u = -y_{\tt c}+v, \qquad u_{\tt c} = y.
\end{equation*} 
Therefore, the closed-loop storage function
\begin{equation*}
 S_{\tt cl}(x,x_{\tt c}) = S(x)+S_{\tt c}(x_{\tt c})
\end{equation*} 
satisfies
\begin{equation*}
 \dot{S}_{\tt cl} = \dot{S}+\dot{S}_{\tt c} \leq y^{\top}u + y^{\top}_{\tt c}u_{\tt c} = 
y^{\top}v-y^{\top}y_{\tt c}+y_{\tt c}^{\top}y = y^{\top}v. 
\end{equation*} 
Consequently, by setting $v=\mathbf{0}$ we guarantee that $S_{\tt cl}(x,x_{\tt c})$ is 
non-increasing, i.e., $\dot{S}_{\tt cl}\leq 0$. Nevertheless, to stabilize a desired equilibrium 
$x^{\star}$, it is necessary to shape the energy of the plant---i.e., $S(x)$---via the controller. 
To 
this end, $x_{\tt c}$ is related to the plant's state through the generation of maps $C: 
\mathcal{X} 
\to \rea^{m}$ such that the sets
$$
\mathcal{M}_\kappa:=\{(x,x_{\tt c}) \in \mathcal{X} \times \rea^{m}\mid C(x)-x_{\tt 
c}=\kappa,\;\kappa \in 
\rea^{m}\},
$$
are invariant. Notice that
\begin{equation*}
 \dot{C}-\dot{x}_{\tt c} = \mathbf{0}, \qquad \forall (x,x_{\tt c})\in\mathcal{M}_{\kappa}.
\end{equation*}
Thus, for any differentiable function $\MAP{\Phi}{\rea^m}{}$, the new storage function
\begin{equation*}
 V(x,x_{\tt c}) := S(x) + S_{\tt c}(x_{\tt c}) + \Phi(C(x)-x_{\tt c}),
\end{equation*} 
satisfies
\begin{equation*}
 \dot{V} = \dot{S}+\dot{S}_{\tt c} + \dot{\Phi} = \dot{S}+\dot{S}_{\tt c} \leq y^{\top}v  
\end{equation*} 
because
\begin{equation*}
 \dot{\Phi} = \nabla^{\top}\Phi \left( \dot{C}-\dot{x}_{\tt c} \right) = 0.
\end{equation*}
Hence, by setting $v=\mathbf{0}$, the storage function $V(x,x_{\tt c})$ is non-increasing. 
Consequently, the stabilization of the desired equilibrium $x^{\star}\in\mathcal{E}$ is achieved if 
$\Phi(C(x)-x_{\tt c})$ and $S_{\tt c}(x_{\tt c})$ can be designed such that $V(x,x_{\tt c})$ is 
locally positive 
definite at $(x^{\star},x_{\tt c}^{\star})$, where $x_{\tt c}^{\star}$ is a control parameter.

\subsection{Asymptotic Stabilization and the Dissipation Obstacle in Control by 
Interconnection}

The design of the functions $\Phi(C(x)-x_{\tt c})$ and $S_{\tt c}(x_{\tt c})$ constitutes the energy
shaping step of this PBC approach. However, given that $\calm_\kappa$ is invariant, achieving 
asymptotic stabilization of the desired equilibrium requires that the initial conditions of the 
plant and controller are contained in the following set:
$$
\Omega_{\kappa}=\{(x,x_{\tt c}) \in \mathcal{X}\times\rea^{m}\mid C(x)-x_{\tt c}=C(x^\star  
)-x_{\tt c}^\star  \}.
$$
In other words, the trajectory should start on the element of $\calm_\kappa$ that contains the 
desired equilibrium, which fixes the initial conditions of the controller. Invoking Sard's 
theorem\footnote{Also known as Morse-Sard theorem.}
\citep{SPIV}, we conclude that $\Omega_{\kappa}$ is a nowhere-dense set. Hence, the 
asymptotic stability property is non-robust \citep{ORTajc20}. In \cite{CASetal09}, the authors 
propose two solutions to this problem: to estimate 
the constant $\kappa^\star:=C(x^{\star})-x^\star_{\tt c}$ or destroy the invariance of 
$\calm_\kappa$ by 
injecting damping.

The CbI approach exposed above depends on the input-output pair. As a consequence, a 
well-documented problem---see, for instance, \cite{ORTetaltac08,MENetal}---occurs when the output 
of the plant to be controlled has relative degree one. In this case, the CbI design might be 
hampered by the dissipation obstacle---see Section \ref{sec:obstacle}. A solution to this problem 
consists in identifying passive outputs with relative degree zero for the 
plant \citep{ORTetaltac08,VENVAN,MENetal}.

\subsection{Other Control by Interconnection Approaches and Extensions}

In a CbI approach different from the one discussed above, the dynamics of the controller are 
allowed to depend on the plant's state---see for instance, \cite{ORTBOR}. This formulation has been 
less explored than the one exposed above but has proven helpful solving specific control problems, 
e.g., injecting damping without measuring the passive output, as done in 
\cite{AIL}.\footnote{Although the PBC interpretation of the result is given in \cite{ORTbook}.} 
Other extensions of CbI include state-dependent interconnections \citep{ORTetaltac08,BORint}, more 
elaborated control dynamics \citep{IOHD,BORint}, and infinite-dimensional systems 
\citep{macchelli15,macchelli16,TOL}. Other aspects that have been studied in the literature are 
the possible invariant sets that can be constructed for a given plant and control structure 
\citep{CER} and the relation between CbI and other PBC techniques \citep{ORTetaltac08,MENBOORT}.

%%%%%%%%%%%%%%%%%%%%%%%
\section{PID Passivity-based Control}\label{sec:PID}
%%%%%%%%%%%%%%%%%%%%%%%
At the beginning of the 21st century, over 90\% of the existing controllers had a PID architecture 
\citep{aastrom}. Nowadays, PID regulators are still the most common and popular controllers in 
engineering applications. These controllers are exceptionally successful at driving a given output 
signal to a constant value. However, PIDs also exhibit fundamental drawbacks. For example, their 
functionality hinges upon the correct selection of their gains, which are often chosen by 
considering a linear behavior of the system to be controlled. If the plant is nonlinear, the typical 
approach is linearizing it around a constant operating point to tune the PID gains. Unfortunately, 
if the plant is highly nonlinear or the range of operation is large, the chosen gains might not be 
appropriate to achieve the control task and could even lead to unstable behaviors. Another 
significant drawback of traditional PID controllers is that, in some applications, the control task 
is not achieved only by driving the output signal to a constant value. Instead, it is required to 
guarantee the stability of a desired equilibrium for the closed-loop system, often done by providing 
a Lyapunov function. PID-PBC offers a solution to the aforementioned problems. In particular, the 
proportional gain is only required to be positive---as it represents the damping injection---and the 
other two gains are chosen to satisfy \eqref{posdef}, trivializing the tuning of the control gains 
for stabilization purposes. Moreover, under these circumstances, the closed-loop storage function is 
a suitable Lyapunov function to prove the stability of the desired equilibrium. Below, we revisit 
the main aspects of the PID-PBC technique.

\subsection{PID Controllers Are Input Strictly Passive}

In the PID-PBC approach, we consider a PID controller described by
\begequ
\begarr{rcl}
\dot x_{\tt c} & = & y \\
u & = & -K_{\tt P} y - K_{\tt I} x_{\tt c} - K_{\tt D} \dot y
\endarr
 \lab{pid}
\endequ
where $y$ corresponds to the passive output of \eqref{sys} and the PID gains $K_{\tt P},K_{\tt 
I},K_{\tt D} \in \rea^{m \times m}$ satisfy
\begin{equation*}
 K_{\tt P}\succ 0, \qquad K_{\tt I}\succ0, \qquad K_{\tt D}\succeq 0.
\end{equation*}
Notably, the system \eqref{pid} is input strictly passive with input-ouput pair $(y,-u)$ and 
storage function given by
\begin{equation*}
 S_{\tt pid}(x_{\tt c},y) = \frac{1}{2}\lVert y \rVert^{2}_{K_{\tt D}} + 
\frac{1}{2}\lVert x_{\tt c}\rVert^{2}_{K_{\tt I}}.
\end{equation*}
To prove such a claim, note that
\begin{equation*}
 \dot{S}_{\tt pid} = y^{\top}K_{\tt D}\dot{y} + x_{\tt c}^{\top}K_{\tt I}\dot{x}_{\tt c} = 
-y^{\top}\left(K_{\tt I}x_{\tt c}+K_{\tt P}y+u \right) + x_{\tt c}^{\top}K_{\tt I}y = -\lVert y 
\rVert^{2}_{K_{\tt P}} - 
y^{\top}u \leq -k_{\tt p}\lVert y \rVert^{2}- y^{\top}u,
\end{equation*}
where $k_{\tt p}$ denotes the minimum eigenvalue of $K_{\tt P}$.

\subsection{Control of Passive Systems With PIDs}

The negative feedback interconnection of \eqref{sys} with \eqref{pid} yields a new passive 
system---see Figure \ref{fig:feed}, where $\Sigma_{1}$ corresponds to the plant and $\Sigma_{2}$ to 
the PID controller \eqref{pid}. Moreover, the closed-loop storage function is given by
\begin{equation*}
 S_{\tt d}(x_{\tt c},x) = S(x)+S_{\tt pid}(x_{\tt c},y),
\end{equation*}
where $y$ can be expressed as a function of $x$ through the expression \eqref{ypar}. 
Thus,\footnote{Note that the external signals $v_{1}$ and $v_{2}$ are set to zero.}
\begin{equation*}
 \dot{S}_{\tt d} = \dot{S}+\dot{S}_{\tt pid} \leq y^{\top}u -\lVert y \rVert^{2}_{K_{\tt P}} - 
y^{\top}u = 
-\lVert y \rVert^{2}_{K_{\tt P}} \leq 0.
\end{equation*}
Accordingly, the stability of $x^{\star}\in\mathcal{E}$ is ensured if $S_{\tt pid}(x_{\tt c},y)$ can 
be 
designed such that $S_{\tt d}(x_{\tt c},x)$ is locally positive definite at 
$(x_{\tt c}^{\star},x^{\star})$. 
However, a simple inspection of $S_{\tt d}(x_{\tt c},x)$ shows that
\begin{equation*}
 S_{\tt d}(x_{\tt c},x) = 0 \implies \left\lbrace \begin{array}{rcl}
                                           S(x) &=& 0 \\ y&=& \mathbf{0} \\ x_{\tt c} &=& 
\mathbf{0}.
                                          \end{array}\right.
\end{equation*}
Hence, $S_{\tt pid}(x_{\tt c},y)$ is not suitable for assigning equilibria different from the 
open-loop 
ones. One way to solve this problem is to relate $x_{\tt c}$ and $x$ through an invariant manifold, 
as 
in CbI. Nevertheless, this leads to the issues related to initial conditions and robustness 
discussed in Section \ref{sec:CbI}. An alternative to overcome this problem is to define the 
integral action in terms of the plant's state. As shown in \cite{TACBOR21,PIDbook}, this is 
possible if, given \eqref{sys}, there exists a function $\MAP{\gamma}{\mathcal{X}}{m}$ such that
 \begin{equation}\label{int}
  \dot{\gamma} = \nabla^{\top}\gamma(x)\left( f(x)+g(x)u \right) = y.
 \end{equation}
Conditions to ensure the existence of $\gamma(x)$ are discussed at length in \cite{TACBOR21}.

If the integrability condition \eqref{int} is satisfied, the PID \eqref{pid} and $S_{\tt 
pid}(x_{\tt c},x)$ can be rewritten as follows:
\begin{eqnarray}
 u &=& -K_{\tt P}y - K_{\tt I}(\gamma(x)+\kappa^{\star})-K_{\tt D}\dot{y} \label{pidnew}\\
\nonumber S_{\tt pid}(x) &=& \dfrac{1}{2}\lVert y \rVert^{2}_{K_{\tt D}} + \frac{1}{2}\lVert 
\gamma(x)+\kappa^{\star} 
\rVert^{2}_{K_{\tt I}}, 
\end{eqnarray}
where
\begin{equation*}
 \kappa^{\star}:=-\left( \gamma^{\star}+K_{\tt I}^{-1}u^{\star} \right).
\end{equation*}
Note that the new closed-loop storage function $S_{\tt d}(x)$ takes the form
\begin{equation*}
 S_{\tt d}(x) = S(x) + \dfrac{1}{2}\lVert y \rVert^{2}_{K_{\tt D}} + \frac{1}{2}\lVert 
\gamma(x)+\kappa^{\star} \rVert^{2}_{K_{\tt I}}.
\end{equation*}
Hence,
\begin{equation}\label{dSdpid}
 \dot{S}_{\tt d} = \dot{S}+y^{\top}K_{\tt D}\dot{y}+\dot{\gamma}^{\top}K_{\tt I}\left( 
\gamma(x)+\kappa^{\star} 
\right)\leq y^{\top}u-y^{\top}u - \lVert y \rVert^{2}_{K_{\tt P}}  = -\lVert y 
\rVert^{2}_{K_{\tt P}}\leq 0.
\end{equation}
Consequently, the stabilization problem reduces to design $K_{\tt I}$ and $K_{\tt D}$ such that 
$S_{\tt d}(x)$ 
is locally positive definite at $x^{\star}$, i.e., \eqref{posdef} holds. Note that this, in 
combination with \eqref{dSdpid}, implies that $S_{\tt d}(x)$ qualifies as a Lyapunov function to 
prove the stability of the desired equilibrium. Moreover, because
\begin{equation*}
 \dot{S}_{\tt d}\equiv 0 \implies y=\mathbf{0}
\end{equation*}
the desired equilibrium is asymptotically stable if
\begin{equation*}
 y=\mathbf{0} \implies x = x^{\star}.
\end{equation*}

\subsection{Well-posedness Conditions}

To guarantee that the PID controller given in \eqref{pidnew} is well-defined and can be implemented, $u$ must have no singularities and be computable without differentiation. These conditions restrict the outputs that can be used to propose the derivative term to relative-degree-one passive outputs. Otherwise, the term $\dot{y}$ implies differentiation of $u$. Additionally, even for relative-degree-one passive outputs---i.e., $y=h(x)$---the derivative term might produce singularities in the control law because
\begin{equation*}
 K_{\tt D}\dot{y} = K_{\tt D}\nabla^{\top}h(x)(f(x)+g(x)u)
\end{equation*}
implies
\begin{equation*}
 \begin{array}{rl}
 & u = -K_{\tt P}h(x)-K_{\tt I}(\gamma(x)+\kappa^{\star})-K_{\tt D}\nabla^{\top}h(x)(f(x)+g(x)u) \\ 
\implies & u = 
-[I+K_{\tt D}\nabla^{\top}h(x)g(x)]^{-1}\left[K_{\tt 
P}h(x)+K_{\tt I}(\gamma(x)+\kappa^{\star})+K_{\tt D}\nabla^{\top}h(x)f(x) \right]
 \end{array}
\end{equation*}
Accordingly, if \eqref{sys} has relative degree one, the well-posedness condition
\begin{equation*}
 \det\left\lbrace I+K_{\tt D}\nabla^{\top}h(x)g(x) \right\rbrace\neq 0
\end{equation*}
must be satisfied to implement the PID passivity-based controller given in \eqref{pidnew}.

As in the CbI case, the dissipation obstacle affects PID passivity-based controllers constructed 
with relative-degree-one passive outputs. To deal with this problem, the integral term must be 
constructed with a relative-degree-zero passive output. Nonetheless, completing a PID scheme with 
these output is not straightforward. In particular, as explained above, $K_{\tt D}$ must be set as 
zero to avoid differentiation of $u$. Furthermore,
the proportional term takes the form
\begin{equation*}
 K_{\tt P}y = K_{\tt P}h(x)+K_{\tt P}j(x)u,
\end{equation*}
and
\begin{equation*}
 u = -K_{\tt P}h(x)-K_{\tt P}j(x)u-K_{\tt I}(\gamma(x)+\kappa^{\star}) \implies u = -\left[I+K_{\tt 
P}j(x) \right]^{-1}\left[K_{\tt P}h(x)+K_{\tt I}(\gamma(x)+\kappa^{\star}) \right].
\end{equation*}
Thus, to implement the controller \eqref{pidnew}, with $K_{\tt D}=\mathbf{0}$, the following 
well-posedness condition must hold:
\begin{equation}\label{well2}
 \det\left\lbrace I+K_{\tt P}j(x) \right\rbrace\neq 0.
\end{equation}

\subsection{PID Passivity-based Control With Zero-relative-degree Passive Outputs}

An alternative to avoid the dissipation obstacle and the condition \eqref{well2}, while completing a PID architecture consists in exploiting the passivity property of a purely integral controller. To illustrate this, suppose \eqref{sys} is passive. Hence, invoking Hill-Moylan's theorem---see Section \ref{sec:par}---we get
\begin{equation*}
 \dot{S} = \nabla^{\top}S(x)f(x)+\nabla^{\top}S(x)g(x)u = 
-\lVert\ell(x)\rVert^{2}-2\ell^{\top}(x)w(x)u - u^{\top}w^{\top}(x)w(x)u+y^{\top}u.
\end{equation*}
Suppose there exists $\gamma(x)$ satisfying \eqref{int}. Define the integral controller
\begin{equation}\label{pid0}
 u = -K_{\tt I}(\gamma(x)+\kappa^{\star})+v.
\end{equation}
Hence, the closed-loop dynamics become
\begin{equation*}
 \dot{x} = f(x)-g(x)K_{\tt I}(\gamma(x)+\kappa^{\star})+g(x)v,
\end{equation*}
and the function
\begin{equation*}
 S_{\tt a}(x) = S(x)+\dfrac{1}{2}\lVert \gamma(x)+\kappa^{\star} \rVert^{2}_{K_{\tt I}}
\end{equation*}
satisfies
\begin{equation*}
\begin{array}{rcl}
  \dot{S}_{\tt a} &=& \dot{S}+\dot{x}_{\tt c}^{\top}K_{\tt I}\left( \gamma(x)+\kappa^{\star} 
\right)\\  &=& 
-\lVert\ell(x)\rVert^{2}-2\ell^{\top}(x)w(x)u - 
u^{\top}w^{\top}(x)w(x)u+y^{\top}u+y^{\top}K_{\tt I}\left( \gamma(x)+\kappa^{\star} \right)\\
  &=& -\lVert \ell(x)-K_{\tt I}\left( \gamma(x)+\kappa^{\star} \right)\rVert^{2}+\left[\nabla S(x)+ 
K_{\tt I}\left( \gamma(x)+\kappa^{\star} \right)\right]^{\top}g(x)v\\
  &\leq& \nabla^{\top}S_{\tt a}(x)g(x)v,
\end{array}
\end{equation*}
where the third equality follows from \eqref{ypar} and \eqref{pid0}. Thus, the closed-loop system 
is passive with storage function $S_{\tt a}(x)$ and passive output
\begin{equation*}
 y_{\tt a}:=g^{\top}(x)\nabla S_{\tt a}(x).
\end{equation*}
Note that the closed-loop system has relative degree one. Hence, if the new well-posedness condition
\begin{equation*}
 \det\left\lbrace I+K_{\tt D}\nabla^{\top}y_{\tt a} \right\rbrace\neq 0
\end{equation*}
holds, the proportional and derivative terms can be added to the integral controller by selecting
\begin{equation*}
 v = -K_{\tt P}y_{\tt a}-K_{\tt D}\dot{y}_{\tt a}.
\end{equation*}
Therefore, the closed-loop storage function 
\begin{equation*}
 S_{\tt d}(x) = S_{\tt a}(x) + \dfrac{1}{2}\lVert y_{\tt a} \rVert^{2}_{K_{\tt D}} 
\end{equation*} 
satisfies
\begin{equation*}
 \dot{S}_{\tt d}\leq -\lVert y_{\tt a} \rVert^{2}_{K_{\tt P}}\leq 0. 
\end{equation*} 
Moreover, the complete PID passivity-controller is given by
\begin{equation*}
 u = -K_{\tt P}y_{\tt a}-K_{\tt I}(\gamma(x)+\kappa^{\star})-K_{\tt D}\dot{y}_{\tt a},
\end{equation*}
which must be implemented as follows:
\begin{equation*}
 u = -\left[I+K_{\tt D}\nabla^{\top}y_{\tt a}g(x) \right]^{-1}\left[K_{\tt 
P}y_{\tt a}+K_{\tt I}(\gamma(x)+\kappa^{\star})+K_{\tt D}f(x) \right].
\end{equation*}

\subsection{Discussion on PID Passivity-based Control}

PID-PBC has proven successful for controlling a wide class of physical systems, see, 
e.g., \cite{SANVER, HERetal, TALetal, MEZetal,CISetal,CISetalpow, ARAetal}. 
In contrast to other nonlinear techniques, the PID-PBC design can be carried out without solving 
PDEs despite the complexity of the dynamics describing the behavior of the system. Examples are 
soft robots and compliant devices \citep{gandhi,BORDASAN}, electromechanical systems 
\citep{BORCISORT}, and electrical circuits \citep{CASetalPI}. In PID-PBC, the proportional term 
carries out the damping injection. Similarly, the integral and derivative terms shape the energy of 
the system. Notably, the integral term assigns the equilibrium for the closed-loop system, playing 
an essential role in stabilizing the desired equilibrium. In mechanical systems, PID-PBC components 
have a direct physical interpretation, where the proportional term can be understood as adding a 
damper to the system, while the integral and derivative terms modify the potential and kinetic 
energy of the system, respectively.

Sometimes it is possible to identify multiple storage functions for a system. In such a case, new 
passive outputs can be generated via the linear combination of the storage functions 
\citep[Chapter 5]{PIDbook}. This approach has been adopted to stabilize mechanical systems in several works,
e.g., \cite{DONetal, gandhi, ROMetal}. Another extension to PID-PBC is the use of 
multipliers, which can be employed to generate new integrable passive outputs \citep{PIDbook}.

Two relevant problems occur when the passive output is not the signal we want to regulate and when 
the desired value for the passive output is different from zero. To address the first issue, several 
authors have explored the possibility of adding an integral action to non-passive outputs while 
preserving some stability properties \cite{DONJUN,FERetal}. To address the second problem, the input 
signal to the PID  controller can be defined as the error between the passive output and its desired 
value. Then, it is investigated if the system is passive with respect to the mentioned error. Such a 
property is known as passivity of the incremental model in \citep{JAYetal}, incremental passivity 
\citep{DEPetal} equilibrium independent passivity \citep{HINetal,SIM}, and shifted passivity in 
\cite{MONetal,VAN,kaw23}.

The relation and overlap of PID-PBC with IDA-PBC are studied in \cite{BORCISORT} and 
\cite{TACBOR21}. Similarly, its relation with CbI is discussed in \cite{MENBOORT}. Furthermore, in 
\cite{VAN}, PID control is analyzed from a
different perspective. Namely, assuming that $\dot{y}$ is computable, they show that the closed-loop system can be represented as a pH system with algebraic constraints. However, the stability analysis of these systems
is beyond the scope of this chapter.
%%%%%%%%%%%%%%%%%%%%%%%%%%%%
%%%%%%%%%%%%%%%%%%%%%%%%%%%%
\section{Interconnection and Damping Assignment}\label{sec:IDA}
%%%%%%%%%%%%%%%%%%%%%%%
Since it was first proposed in \cite{IDAaut}, IDA-PBC has established itself as a popular nonlinear 
control technique, which has proven suitable for controlling a broad range of systems---see, for 
example, \cite{PETetal,RAMetal,DONetalrobust,HEetal,CUP,YUK,IPMC,ryalat,FRANCAS}. This nonlinear 
control technique is considered the most general PBC 
strategy \citep{ORTetaltac08,wu} because the set of plants that can be controlled is larger than in 
other PBC approaches. A fundamental difference between IDA-PBC and other passivity approaches is 
that the energy shaping does not hinge upon identifying an appropriate passive output, as in CbI or 
PID-PBC.

\subsection{Stabilization via Interconnection and Damping Assignment Passivity-based Control}

In the IDA-PBC method, we consider the following target dynamics
\begin{equation}\label{IDAclsys}
 \dot{x} = F_{\tt d}(x)\nabla H_{\tt d}(x),
\end{equation}
where $\MAP{H_{\tt d}}{\mathcal{X}}{}$ must be positive definite at the desired 
equilibrium 
$x^{\star}\in\mathcal{E}$ in a neighborhood $\mathcal{D}\subseteq\mathcal{X}$ of $x^{\star}$, 
i.e.,
 \begin{equation}\label{IDApos}
  H_{\tt d}^{\star}=0 \quad \text{and} \quad  H_{\tt d}(x)>0, \; \forall 
x\in\mathcal{D}-\{x^{\star}\},
\end{equation}
and $\MAP{F_{\tt d}}{\mathcal{X}}{n\times n}$ needs to satisfy
\begin{equation}\label{IDAsym}
 F_{\tt d}(x)+F_{\tt d}^{\top}(x)\preceq 0.
\end{equation} 
Note that \eqref{IDAsym} implies that
\begin{equation*}
 \dot{H}_{\tt d} = \nabla^{\top} H_{\tt d}(x)F_{\tt d}(x)\nabla H_{\tt d}(x)= 
\dfrac{1}{2}\nabla^{\top} H_{\tt d}(x)\left[F_{\tt d}(x)+F_{\tt d}^{\top}(x) \right]\nabla H_{\tt 
d}(x)\leq 0.
\end{equation*}
Therefore, if \eqref{IDApos} and \eqref{IDAsym} are satisfied, the function $H_{\tt d}(x)$ qualifies 
as 
a Lyapunov function to prove the (local) stability of
$x^{\star}$. Furthermore, the desired equilibrium is (locally) asymptotically stable if
\begin{equation*}
 \dot{H}_{\tt d} \equiv 0 \implies x = x^{\star}.
\end{equation*}
Additionally, the stability properties of $x^{\star}$ are global if $H_{\tt d}(x)$ is radially
unbounded. 

We stress that the matrix $F_{\tt d}(x)$ and the closed-loop storage function $H_{\tt d}(x)$ 
cannot be 
chosen arbitrarily. This can be noted by equating the dynamics of $x$ given in \eqref{sys} with
\eqref{IDAclsys} to obtain
\begin{equation*}
  F_{\tt d}(x)\nabla H_{\tt d}(x)-f(x) = g(x)u.
\end{equation*}
Premultiplying the equation above by $g^{\perp}(x)$ yields
\begin{equation}\label{match}
 g^{\perp}(x)\left\lbrace F_{\tt d}(x)\nabla H_{\tt d}(x) - f(x)\right\rbrace =
\mathbf{0}.
\end{equation}
The system of PDEs given in \eqref{match} is known as the matching equations \citep{IDAsurvey}. 
Notably, if there exist $F_{\tt d}(x)$ and $H_{\tt d}(x)$ solving \eqref{match}, the controller 
that 
yields the desired closed-loop system \eqref{IDAclsys} is given by
\begin{equation*}
  u = g^{+}(x)\left\lbrace F_{\tt d}(x)\nabla H_{\tt d}(x) -
f(x)\right\rbrace,
\end{equation*}
where $g^{+}(x):=\left[ g^{\top}(x)g(x) \right]^{-1}g^{\top}(x)$.

\subsection{Comments on Interconnection and Damping Assignment 
Passivity-based Control}

Sometimes, $F_{\tt d}(x)$ is represented as $J_{\tt d}(x)-R_{\tt d}(x)$, where $J_{\tt d}(x) = 
-J_{\tt d}^{\top}(x)$ and $R_{\tt d}(x)\succeq 0$. In this representation, $J_{\tt d}(x)$ and 
$R_{\tt d}(x)$ are called the interconnection and damping matrices, respectively---whence, the name 
interconnection and damping assignment.

Finding a solution to \eqref{match}---while ensuring that $F_{\tt d}(x)$ and $H_{\tt d}(x)$ 
satisfy the 
conditions \eqref{IDApos} and \eqref{IDAsym}---represents the main bottleneck in IDA-PBC. To 
address this problem, several references focus on systematic methods to solve such PDEs, e.g., 
\cite{nunna,cieza2018ida,pfaffian}. Unfortunately, finding a general constructive method to solve 
the PDEs arising in IDA-PBC remains an open problem.

IDA-PBC has been particularly successful in solving the 
stabilization problem for underactuated mechanical systems \citep{ORTtac2002,acrobot,DONetalrobust}. 
In this specific scenario, the matching equations are often split into two sets of PDEs, where one 
set corresponds to the potential energy shaping and the other to the kinetic energy shaping. Similar 
to the  general IDA-PBC case, great effort has been put into providing constructive approaches to 
solve the matching equations; see, for example, \cite{acosta,viola,SIDA,cieza2019ida,HAR,ARP}. 
Alternatively, learning approaches have been adopted to solve the matching equations 
\citep{SANBAEZ,siri}. Although IDA-PBC is unaffected by the dissipation obstacle, natural damping 
may complicate solving the matching equations when kinetic energy shaping is required 
\citep{GOM04}. To address this issue, some works focus on compensating friction before applying the 
IDA-PBC method to mechanical systems; see, for instance, \cite{FRANCO}.

Several extensions of IDA-PBC can be found in the literature, including IDA-PBC robust to 
disturbances \citep{DONetalrobust}, adaptive versions of IDA-PBC \citep{POP}, IDA-PBC approaches 
for discrete-time systems \citep{moreschini}, and trajectory-tracking controllers based on IDA-PBC 
\citep{YAG}. Furthermore, we refer the reader to
\cite{ORTetaltac08} and \cite{TACBOR21} for a discussion on the relation of IDA-PBC with 
CbI and PID-PBC, respectively.

%%%%%%%%%%%%%%%%%%%%%%%
\section{Summary}
%%%%%%%%%%%%%%%%%%%%%%%

\bul Passivity is fundamental in many nonlinear dynamical systems, including a broad range of physical systems. This property implies that a system is unable to generate energy.

\bul The negative feedback interconnection of two passive systems yields another passive system. This fact can be exploited for control and stability analysis purposes.

\bul The two main components of PBC techniques are energy shaping and damping injection. These two elements have a close relation to the concept of energy. Hence, they often have a physical interpretation, making PBC intuitive.

\bul In CbI, the plant and controller are passive systems, where the latter is designed to shape the energy of the plant in a desired manner. To this end, invariant functions are often identified to relate the states of both systems.

\bul In PID-PBC, the controller is given by traditional PID architecture with the passive output of the plant used as the input signal to the PID. To this end, the output must be integrable and well-posedness conditions must be met. In contrast to traditional PIDs, the tuning process to ensure stability is transparent in PID-PBC. On the other hand, to ensure good performance it is necessary to do some gain tuning among all possible positive gains,

\bul IDA-PBC is the most general PBC approach. This versatile nonlinear control strategy has proven suitable for stabilizing a broad range of systems. However, its implementation may be hampered by the need to solve PDEs.

%%%%%%%%%%%%%%%%%%%%%%%
\bibliographystyle{plain}
\bibliography{refs_encyclopedia}

\end{document}